\documentclass[aps,APL,twocolumn,showpacs,superscriptaddress,groupedaddress]{revtex4}  
\usepackage{graphicx}  
\usepackage{dcolumn}   
\usepackage{bm}        
\usepackage{amssymb}   
\usepackage{textcomp} 
\usepackage{threeparttable}

\usepackage{setspace}
\usepackage{amsmath}
\usepackage{booktabs}

\begin{document}



\title{Non-polar Flexoelectrooptic Effect in Blue Phase Liquid Crystals}
\date{\today}

\author{B. I. Outram}
\email{benjamin.outram@gmail.com}
\address{Department of Engineering Science, University of Oxford, Parks Road, Oxford OX1 3PJ, United Kingdom}

\author{S. J. Elston}
\address{Department of Engineering Science, University of Oxford, Parks Road, Oxford OX1 3PJ, United Kingdom}

\author{F. Castles}
\address{Centre of Molecular Materials for Photonics and Electronics, Department of Engineering, University of Cambridge,
9 JJ Thomson Avenue, Cambridge CB3 0FA, United Kingdom}

\author{M. M. Qasim}
\address{Centre of Molecular Materials for Photonics and Electronics, Department of Engineering, University of Cambridge,
9 JJ Thomson Avenue, Cambridge CB3 0FA, United Kingdom}

\author{H. Coles}
\address{Centre of Molecular Materials for Photonics and Electronics, Department of Engineering, University of Cambridge,
9 JJ Thomson Avenue, Cambridge CB3 0FA, United Kingdom}

\author{H.-Y. Chen}
\address{Department of Photonics, Feng Chia University, Taichung, Taiwan 40724, Republic of China}

\author{S.-F. Lu}
\address{Department of Photonics, Feng Chia University, Taichung, Taiwan 40724, Republic of China}

\begin{abstract}
{Blue phase liquid crystals are not usually considered to exhibit a flexoelectrooptic effect, due to the polar nature of flexoelectric switching and the cubic or amorphous structure of blue phases.  Here, we derive the form of the flexoelectric contribution to the Kerr constant of blue phases, and experimentally demonstrate and measure the separate contributions to the Kerr constant arising from flexoelectric and dielectric effects.  Hence, a non-polar flexoelectrooptic effect is demonstrated in blue phase liquid crystals, which will have consequences for the engineering of novel blue-phase electrooptic technology.}
\end{abstract}
\maketitle

The properties of chiral liquid crystals make them suited to studying unique linear and quadratic electrooptic phenomena. The chiral-flexoelectrooptic effect results from a coupling between bend and splay curvature distortions and electric polarization.  In the cholesteric phase, the effect produces a linear rotation in the optic axis as a function of electric field strength \cite{Pat87}, however due to a cubic or amorphous structure, flexoelectricity is often presumed to produce no direct field-induced birefringence in blue phase liquid crystals (BPLCs), although flexoelectric-induced structural changes in blue phases have been investigated theoretically \cite{tiribocchi2013flexoelectric,alexander2007flexoelectric}. The possibility of flexoelectrooptic switching in blue phases \emph{via} electrostriction  \cite{kitzerow2010blue}, or due to flexoelectric or order-electric polarization in the vicinity of defects \cite{dmitrienko1989electro} has also been suggested, and a flexoelectrooptic effect may have been observed at high fields \cite{chen2013unusual}.  Kerr effects in the blue phases have been attributed to the direct coupling between the director orientation and electric field, known as the dielectric effect.  However, in this paper we describe how the coupling of flexoelectric polarization to director curvature distortion produces a quadratic flexoelectooptic effect in BPLCs, and derive an expression relating flexoelectric parameters to the Kerr constant. In addition, we measure the independent contributions to the Kerr switching arising from dielectric and flexoelectric contributions in both dielectric- and flexoelectric-dominated BPLCs.  It is demonstrated that both effects contribute to Kerr switching, and that the separate contributions combine constructively or destructively in cases of positive or negative dielectric anisotropy, respectively.  Geometric factors relating liquid crystal parameters to both flexoelectric and dielectric contributions to Kerr switching are determined from measurements, and are found to be similar in magnitude.  Exploitation of this non-polar flexoelectrooptic phenomena thus provides an alternative approach for improving the Kerr behaviour and functionality of blue phase materials for electrooptic technologies, for which advances are of particular present interest.

In the presence of a field $E$, the dielectric energy (proportional to $\Delta\varepsilon(\mathrm{\mathbf{\hat{n}}}\cdot \mathrm{\mathbf{E}})^2$) causes a deflection of the director $\mathrm{\mathbf{\hat{n}}}$ towards the field direction, and therefore increases the refractive index in this direction relative to orthogonal directions, resulting in a positive induced blue phase birefringence $\delta n$ parallel to the field, in the cases of positive dielectric anisotropy $\Delta\varepsilon$.  In general, biaxiality is induced \cite{porsch1984electric},  however the effect is negligible relative to the uniaxial component, and for simplicity we do not consider this effect here.  In the case of a negative $\Delta\varepsilon$, the reverse is true, and the result is a negative induced birefringence.  Gerber derived an approximate form for the Kerr constant \cite{gerber1985electro}, defined by the expression $\delta n=\lambda K E^2$, where $\lambda$ is the wavelength of light, $E$ is an electric field, and $K$ is the Kerr constant of the form
\begin{equation}\label{gerberGd}
K_\mathrm{d}\approx G_\mathrm{d}\left[\frac{\Delta n}{\lambda q^2}\left(\frac{\Delta\varepsilon\varepsilon_0}{K_2}\right)\right],
\end{equation}
where we have included a subscript ``d'' to denote that this is the dielectric contribution, and have added a unitless ``geometric factor'' $G_\mathrm{d}$ to account for the relationship between the physical parameters and the induced birefringence, which is dependent on the blue phase structure and may depend on the type of blue phase (PBI, BPII or BPIII), which for simplicity, we also do not consider here.  $\Delta n = n_\mathrm{e}-n_\mathrm{o}$, where $n_\mathrm{o}$ and $n_\mathrm{e}$ are the local ordinary and extraordinary liquid crystal refractive indices, $K_2$ is the twist elastic constant, and $q=2\pi/P$, where the pitch $P$ is assumed to be equal to the blue phase lattice parameter \cite{miller1996lattice}.

In the case of a cholesteric liquid crystal in which the pitch $P\ll\lambda$ and in which $\Delta n \ll n_\mathrm{o}$, one can approximate the optics as being uniaxial with an effective birefringence $\Delta n'\approx-\Delta n/2$, with the optic axis parallel to the helicoidal axis $\mathrm{\mathbf{\hat{n}}}'$.  In the presence of a field $\mathrm{\mathbf{E}}=E_z\mathrm{\mathbf{\hat{z}}}$ at an angle $\theta$ to $\mathrm{\mathbf{\hat{n}}}'$, flexoelectricity causes a rotation of the optic axis about a field component that is perpendicular to $\mathrm{\mathbf{\hat{n}}}'$, whose magnitude is $E_\perp=E\sin\theta$, where $\cos\theta = \mathrm{\mathbf{\hat{n}}}'\cdot \mathrm{\mathbf{\hat{z}}}$.  The $z$-component of the cholesteric's refractive index within the field can, for small $\Delta n$ and $\xi$, be approximately expressed
\begin{equation}\label{eq:indicatrixcholesteric}
n_\theta\approx n_\mathrm{o}'\left(1+\cos^2\theta(1-\xi^2\sin^2\theta)\frac{\Delta n'}{n_\mathrm{o}'}\right)
\end{equation}
where $n_\mathrm{o}'$ is the effective refractive index perpendicular to the helicoidal axis and
\begin{equation}
\xi = \frac{e_1-e_3}{K_1+K_3}\frac{E}{q}.
\end{equation}
where $e_1-e_3$ is the difference between the flexoelectric parameters as defined by Meyer \cite{mey69}, and $K_1$ and $K_3$ are the splay and bend elastic constants.

The cubic blue phases consist of an array of so-called ``double-twist'' cylinders in which the director orientation rotates about a radial direction from the cylinder center \cite{wright1989crystalline}.  Heuristically, one can imagine the blue phase as being analogous to a cholesteric, but in which the helicoidal axis has components in multiple directions.  To estimate the $z$-component of the effective refractive index of the blue phase, we sum the optical relative permittivity of the cholesteric ($n_\theta^2$ from Eq.~\ref{eq:indicatrixcholesteric}) over all angles of $\theta$ and $\phi$, where $\phi$ is the angle about the $z$-axis, and normalize. Finally, by considering that for small distortion $\sum\nolimits_{i=x,y,z}\varepsilon_i\approx \mathrm{constant}$, the effective induced birefringence due to flexoelectricity of the blue phase can be found to be
\begin{equation}\label{eq:outram}
\delta n_\mathrm{f} \approx \frac{1}{10}\Delta n\xi^2
\end{equation}
where the factor of $1/10$ is an effective flexoelectric geometric factor, is in part due to the approximations that have been made in the derivation, and is not necessarily expected to correspond precisely to the experimentally determined value.  Thus, the contribution to the Kerr constant $K$ due to flexoelectricity can be expressed
\begin{equation}\label{eq:outram1}
K_\mathrm{f}=G_\mathrm{f}\left[\frac{\Delta n}{\lambda q^2}\left(\frac{e_1-e_3}{K_1+K_3}\right)^2\right].
\end{equation}
where $G_\mathrm{f}$ is the flexoelectric geometric factor.

\begin{figure}
\begin{center}
\includegraphics[width=0.45\textwidth]{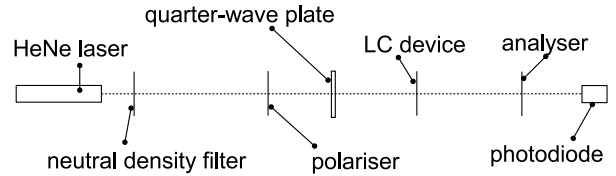}
\end{center}
\vspace{-15mm}
\caption{A schematic of the experimental arrangement.  A circular polarization is generated using a linear polarizer and quarter-wave plate.  A liquid crystal cell with IPS-type electrodes is followed by an analyser set at 45 degrees to the applied electric field direction, and a photo-diode detects the transmitted intensity.}\label{fig:exparr}
\end{figure}

The Kerr behaviour is investigated using the experimental arrangement shown in figure \ref{fig:exparr}.  A liquid crystal device, of thickness $d=5$ \textmu m, with a field applied in the plane of the device is placed between crossed polarizers whose axes are set at 45$^\circ$ to the applied field direction. A circular incident polarization is generated using a quarter-wave plate.  The transmission of the device within this arrangement is determined using a laser ($\lambda=633$ nm) and photo-diode detector, and for small birefringence is given by
\begin{equation}\label{eq:5Tkerr2}
T=\frac{1}{2}+\frac{\pi\delta n d}{\lambda}
\end{equation}
where $d$ is the cell thickness.

To measure $K_\mathrm{d}$ and $K_\mathrm{f}$ we exploit the fact that the flexoelectric switching, whose underlying distortion depends on the field polarity, is suppressed at a high enough frequency, while the dielectric distortion is not suppressed, even at much higher frequencies (but at frequencies lower than the relaxation in dielectric permittivity due to the mobility of molecular dipole reorientations) \cite{outram2013frequency}. The frequency at which flexoelectric switching is suppressed in the cholesteric phase, given by 
\begin{equation}\label{eq:fcrit}
f_\mathrm{flexo}\approx\frac{q^2}{4\pi\gamma}(K_1+K_3),
\end{equation}
where $\gamma$ is a viscosity \cite{RudThes}, is expected to be similar to the frequency at which the flexoelectric contribution to Kerr switching in blue phases is suppressed.  We may therefore assume that the flexoelectric Kerr contribution is frequency dependent, and that the total blue phase Kerr constant can be expressed,
\begin{equation}\label{eq:5kerr1}
K(\omega) = \frac{K_\mathrm{f}}{1+\omega^2\tau_\mathrm{f}^2} + K_\mathrm{d}
\end{equation}
where $\tau_f = 1/(2\pi f_\mathrm{flexo})$.  $K_\mathrm{d}$ and $K_\mathrm{f}$ can be determined by measuring the birefringence as a function of frequency using an amplitude-modulated (AM) carrier signal whose frequency we vary, and observing a relaxation in the birefringence due to the suppression in flexoelectrooptic switching at high frequencies.

\begin{figure}
\begin{center}
\includegraphics[width=0.45\textwidth]{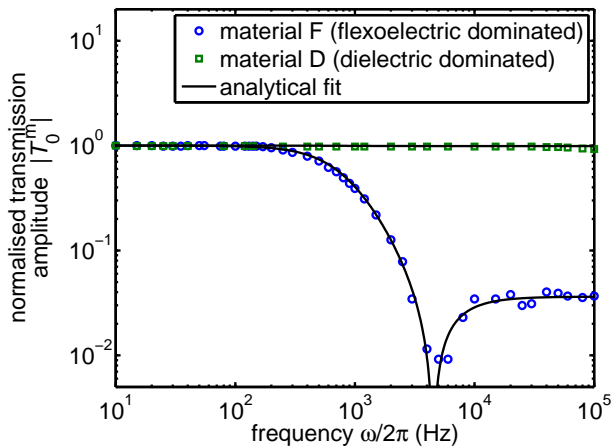}
\end{center}
\caption{The normalised transmission amplitude at the modulation frequency, $|T_0^\mathrm{m}|$, is shown as a function of the driving frequency $\omega$, under the application of a field of the form $E=E_0\sin(\omega t)\sin^2\left(\frac{\omega_\mathrm{m}}{2}t\right)$, in both flexoelectric and dielectric dominated BPLC materials, on a log-log scale.  The analytical model is of the form given in Eq.~\ref{eq:Kdominant} (solid line).  A point at which the flexoelectric and dielectric contributions combine to give $K=0$, and therefore at which $T_0^\mathrm{m}=0$, is evident in the model at between 4 and 5 kHz.}\label{fig:spec}
\end{figure}

 \begin{table*}
\begin{threeparttable}[b]
\begin{center}
\caption{Physical parameters of BPLC mixtures F and D.   The error in $K_{\omega\rightarrow 0}$ is estimated to be 10\%, with the dominant sources of error being uncertainty in device thickness and field strength.  Unless otherwise specified, measurements were taken at $T =$ 50$^\circ$C for mixture F and 34$^\circ$C for mixture D, which corresponds to a reduced temperature (defined as $T_{\mathrm{NI}}-T$, where $T_{\mathrm{NI}}$ is the BP-isotropic transition temperature) of approximately 2$^\circ$C for both materials,}\label{tab:params}
\vspace{5mm}
\begin{ruledtabular}
\begin{tabular}{@{}lccccccr}

										&		 $\Delta\varepsilon$ & $K_2$ (pN)	&		$\frac{e_1-e_3}{K_1+K_3}$	(CN$^{-1}$m$^{-1}$)	&	$\kappa$											&	$\gamma$	(Pa$\cdot$s)			&	$\tau$		(\textmu s)					&	$K_{\omega\rightarrow 0}$ (nmV$^{-2}$)	\\
\hline
 material F					& 	$-0.083$\tnote{a}		& $2$ 					& $1.1$ 													& $-3\times 10^{-1}$							& $\approx 0.6$		& $\tau_\mathrm{f}=190$ 		& $0.022$	\\
 material D					&  	1.6							& 2 						&  	$0.011$												&  	$>6\times 10^{4}$						&  	$\approx 0.1$		&		$\tau_\mathrm{d} = 23$	&		$0.13$\\

\end{tabular}
\begin{tablenotes}
\item[a] Measured at $T=40^\circ$C.
\end{tablenotes}
\end{ruledtabular}

\end{center}
\end{threeparttable}
\end{table*}

Consider a field of the form
\begin{equation}\label{eq:appfield}
E=E_0\sin(\omega t)\sin^2\left(\frac{\omega_\mathrm{m}}{2}t\right),
\end{equation}
in which $\omega$ is a test angular frequency, and $\omega_\mathrm{m}$ is a fixed AM frequency where $\omega_\mathrm{m} \ll 2\pi f_\mathrm{flexoelectric}$.   Substituting Eqs.~\ref{eq:5kerr1} and \ref{eq:appfield} into the expression $\delta n=\lambda K E^2$, and then substituting the resultant form of $\delta n$ into Eq.~\ref{eq:5Tkerr2}, we find that 
\begin{equation}\label{eq:Kdominant}
T_0^\mathrm{m}=\frac{\pi d}{4}\left(\frac{K_\mathrm{f}}{1+\omega^2\tau_\mathrm{f}^2} + K_\mathrm{d}\right)E_0^2
\end{equation}
where $T_0^\mathrm{m}$ is the amplitude of the transmission at the AM frequency $\omega_\mathrm{m}$.  Figure \ref{fig:spec} shows normalised, experimentally-determined $|T_0^\mathrm{m}|$ as a function of the driving frequency $\omega$, for cells with interdigitated electrodes ($d=5$ \textmu m, with $9$ \textmu m electrode gap and $5$ \textmu m wide electrodes) filled with two blue phase materials: a flexoelectric-dominated material, mixture F, comprising chiral dopant 4.2 wt\% BDH1281 \cite{coles2009liquid}, and 24 wt\% each of FFO5OFF, FFO7OFF, FFO9OFF and FFO11OFF \cite{coles2005liquid} (provided by the University of Cambridge); and a dielectric-dominated material, mixture D comprising nematic 79 wt\% FCU-LCM10 and 21 wt\% chiral dopant FCU-NYCL (provided by the University of Feng Chia) \cite{chen2013unusual}. The relative size of flexoelectric and dielectric field-induced distortion in the two materials can be quantified using the dimensionless ratio between Eqs. \ref{gerberGd} and \ref{eq:outram1}, but ignoring the geometric factors which relate the distortions to the Kerr constant, which is given by
\begin{equation}\label{eq:kappa}
\kappa=\left. \left\{\frac{\Delta\varepsilon\varepsilon_0}{K_2}\right\} \middle/ \left\{\frac{e_1-e_3}{K_1+K_3}\right\}^2\right. .
\end{equation}
Material D has a large $\kappa$ of $>6\times10^{4}$, while material F has a small and negative $\kappa$ of approximately $-3\times10^{-1}$.  To determine values of $\kappa$, $\Delta\varepsilon$ was determined from capacitance measurements of liquid crystal filled devices while in the cholesteric phase. $(e_1-e_3)/(K_1+K_3)$ in mixture D was determined using a modified version of the method in reference \cite{outram2012flexoelectric} (which also allowed the measurement of $K_2$) and in mixture F using a standard uniform-lying-helix method \cite{castles2012flexoelectric}. $K_2$ for mixture F was estimated based on values reported for similar compounds \cite{jin1998highly}. $\gamma$ was determined from the relation $2\gamma=(K_1+K_3)q^2\tau_\mathrm{f}$ (c.f. Eq.~\ref{eq:fcrit}) for mixture F, where $\tau_\mathrm{f}$ was determined using the analytical fit shown in figure \ref{fig:spec}.  $\gamma$ for mixture D was determined using a related expression, $\gamma=K_2q^2\tau_\mathrm{d}$, where $\tau_\mathrm{d}$ is the characteristic time of the dielectric distortion, and was measured using the transmission of the blue phase device within the experimental arrangement in figure~\ref{fig:exparr} under the application of bursts of 200 kHz AC.  The total low-frequency Kerr constant, $K_{\omega\rightarrow 0}$ was measured by analysing transmission of devices also within the experimental arrangement in figure~\ref{fig:exparr}. These liquid crystal parameters used to calculate $\kappa$ are given in table \ref{tab:params}.  

From Eq.~\ref{eq:5kerr1}, it can be seen that at low frequency, $K=K_\mathrm{f}+K_\mathrm{d}$, and for mixture F, the flexoelectric component is dominant and results in a positive $K$. At high frequency $K=K_\mathrm{d}$, and mixture F's negative $\Delta\varepsilon$ results in a negative $K$ at high frequency.    There is a  relaxation in $T_0^\mathrm{m}$, shown in figure~\ref{fig:spec}, at a frequency consistent with Eq. \ref{eq:fcrit}, and therefore consistent with a relaxation in the flexoelectric switching of mixture~F.  The relaxation in $T_0^\mathrm{m}$ occurs at a large enough frequency that the effect of ionic impurities can be ruled out as being the source of the observed relaxation.  Further, other than a relaxation due to the suppression of the polar flexoelectric switching, dielectric measurements of mixture F when in the cholesteric phase at cooler temperatures show no further relaxations in dielectric properties within the range of frequencies used in this study, which rules out other dielectric dispersion phenomena.   $T_0^\mathrm{m}$ passes through zero when the dielectric and flexoelectric Kerr contributions exactly cancel out. Thus, $K_\mathrm{f}+K_\mathrm{d}$ combine destructively in the case of negative $\Delta\varepsilon$.   In contrast, for mixture D, having negligible flexoelectric switching and a positive $\Delta\varepsilon$, $T_0^\mathrm{m}$ shows no suppression as a function of frequency, and the mixture has a positive, frequency-independent Kerr constant in this frequency range.  For mixture F, a relaxation model of the form given in Eq. \ref{eq:Kdominant} is used to determine $K_\mathrm{f}/K_\mathrm{d}$ and $\tau_f$.  For mixture D, for which $T_0^\mathrm{m}$ does not show a relaxation, the model is used to provide an upper bound on $K_\mathrm{f}/K_\mathrm{d}$ within the experimental error.

\begin{table}
\caption{Experimentally determined ratios of contributions to the Kerr switching due to flexoelectric and dielectric effects in blue phase materials D and F.  The values of the flexoelectric and dielectric contributions, $K_\mathrm{f}$ and $K_\mathrm{d}$ are calculated by substituting $K_\mathrm{f}/K_\mathrm{d}$ and the total Kerr constant $K_{\omega\rightarrow 0}$ (given in table \ref{tab:params}) into Eq.~\ref{eq:5kerr1} in the limit $\omega\rightarrow 0$.  The error in $K_\mathrm{f}/K_\mathrm{d}$ for material F is much less than that in $K_\mathrm{d}$ and $K_\mathrm{f}$, which have an estimated error of 10\% due to uncertainty in $K_{\omega\rightarrow 0}$ (see table \ref{tab:params}).}\label{tab:1}
\vspace{5mm}
\begin{ruledtabular}
\begin{tabular}{@{}lccr}
										&$K_\mathrm{f}/K_\mathrm{d}$			&		$K_\mathrm{d}$ 	(nmV$^{-2}$)										& $K_\mathrm{f}$ (nmV$^{-2}$)				\vspace{0mm}\\
\hline 
	material F				& $-$31																						&   $-$0.00074													&0.023	\\
	material D				&	$<$0.01																				&			0.13																				&	$<$0.001\\
\end{tabular}
\end{ruledtabular}
\end{table}

Independent measurement of the total Kerr constant at low frequency, $K_{\omega\rightarrow 0}$, allows the determination of the values of $K_\mathrm{d}$ and $K_\mathrm{f}$, which are summarised in table~\ref{tab:1}. The geometric factors can then be determined by substituting the independently measured physical parameters relating to dielectric and flexoelectric distortions, given in table \ref{tab:params}, into Eqs. \ref{gerberGd} and \ref{eq:outram1}.  The experimental value of $K_\mathrm{d}$ from material D, and $K_\mathrm{f}$ from material F, are then used to determine $G_\mathrm{d}$ and $G_\mathrm{f}$ respectively.  We find that both $G_\mathrm{d} = 0.11 \pm 0.02$ and $G_\mathrm{f}=0.11 \pm 0.02$.  The error is primarily due to uncertainty in the liquid crystal physical parameters  and these factors may depend on the specific type of blue phase present.   Both flexoelectric and dielectric contributions to $K$ are similar in magnitude.  It should be noted, however, that for typical liquid crystal materials, $\kappa \gg 1$ and in this case the dielectric contribution will dominate.

While we have considered flexoelectric and dielectric contributions to the blue phase Kerr effect, we note that the structure of the blue phases, including a network of defects and resultant gradients in field and order parameter, mean that both gradient flexoelectricity (dependent on $e_1+e_3)$  and order electricity \cite{durand1990order} potentially also contribute.  Although it is difficult to decouple these effects in an electrooptic study, the fact that non-polar local distortions can contribute to quadratic electrooptic effects has been demonstrated and can now inform material development and numerical modelling of blue phase systems.  The geometric factors may also vary between the different blue phases, BPI, PBII or PBIII, which this  initial  study  has not differentiated between. In addition, a polar electrooptic effect has been noted elsewhere \cite{flynn2012polar}.

The non-polar flexoelectrooptic effect described and demonstrated in this work has a similar influence on the magnitude of the Kerr effect in blue phase systems as the dielectric effect, as demonstrated by the similar magnitudes of the experimentally determined geometric factors relating physical parameters to relevant Kerr constants.  Large-Kerr-constant blue phase materials for emerging electro-optic technologies, including potential field-sequential colour and stereoscopic displays and projectors \cite{kikuchi2002polymer,he2012fast}, and liquid-crystal-over-silicon (LCOS) devices including adaptive optics and holographic projectors \cite{chen2009high,wilkinson2006phase}, may therefore be developed that exploit the flexoelectric properties of liquid crystals. Furthermore, the electro-optic technique developed in this work to determine the individual flexoelectric and dielectric contributions to the Kerr constant has shown that the contributions combine constructively in the case where the dielectric anisotropy is positive, and destructively in the case of negative dielectric anisotropy.

\vspace{18pt}
\textbf{Acknowledgements }-- B.I.O. wishes to thank the EPSRC and Merck Chemicals Ltd, UK, a subsidiary of Merck KGaA, Darmstadt, Germany, for financial support through a CASE award.


\end{document}